**Type of Article: Research Article/Original Article**

**Title:** Face morphometric profiles of groups as early markers for certain diseases?

**Authors:** Roberto Herrero[1], Yoanna Martinez-Diaz[2], Heydi Mendez-Vazquez[2], Joan Nieves[1] and Augusto Gonzalez[1]

[1]Instituto de Cibernética, Matemática y Física, Vedado, La Habana, Cuba

[2]Centro de Aplicaciones de Tecnologías Avanzadas, Miramar, La Habana

**Corresponding author:** Roberto Herrero, Instituto de Cibernética, Matemática y Física. La Habana, Cuba. Tel: +5358277440, Fax:-, E-mail: rohepe@icimaf.cu

**Abstract:** Purpose: To show preliminary results on the face morphometry profile of the Cuban population and to argue that it could be used to define early markers for diseases such as Alzheimer's or cancer. Methods: A dataset composed of photos of 200000 men is processed. Facial landmarks are extracted by means of the DLIB library and distances between them are computed. By clustering samples with similar facial traits, groups are formed and their densities inside the population are computed. Results: The face morphometry profiles for two age cohorts are obtained, showing the population dynamics. Genes involved in facial development are shown to be related to Alzheimer's disease and cancer. Conclusions: Late multifactorial diseases develop against the genetic background of each individual, which is expressed by its face morphometry. The latter can be thus considered a risk factor.

**Abreviations:** AD: Alzheimer's disease; FM: Face morphometry; PCA: Principal Component Analysis; TCGA: The Cancer Genome Atlas

## 1. Introduction.

Population-wide studies, such as infant height and weight tables **[1]**, are useful tools for early detection of metabolic disorders or any other kind of disruption of normal state in children. By comparing the infant's parameters with the reference values, we get a first signal of a possible disorder and may proceed to further studies. At present, when the molecular approach to medicine is prevailing more and more in clinical analyses **[2]**, genetic and epigenetic markers **[3]** and their reference values in a population become extremely important in many diseases.

However, genetic studies are still relatively expensive. In certain situations, in particular the "trivial" Down's syndrome, a direct examination of the face is enough for diagnosis. Less known, but similarly strong as a marker, is the diagnosis of Down's syndrome from fingerprints patterns **[4]**. Both markers: face characteristics and fingerprints, are easily obtained. In a sense, they are indirect markers, integrating genetic an epigenetic information. As an example of paper using face morphometry (FM) for the diagnosis of a set of syndromes we may cite Ref. **[5]**.

Instead of pretending a diagnosis from indirect markers, we may try to define risk groups inside the population. It is well known that, in many diseases, the risk has an ethnic or group component **[6]**. Consider, for example, the following cancer risk data from 423 cancer registries in the world (**Ref.[7]**, Supplementary information). A Principal Component Analysis (PCA) **[8]** of the data shows that ethnic and cultural groups exhibit distinct patterns.

For a given population, in particular the Cuban population, mixing between groups could be an element hindering ethnic origins or other factors. A more detailed analysis based on FM measurements, for example, could be a valuable tool to identify groups inside the population and, indirectly, multivariate markers for predisposition to certain diseases.

## 2. Results

### 2.1. Face morphometric profile of the Cuban population.

We performed a preliminary analysis of FM data of the Cuban population. Two cohorts of men, each with 100,000 randomly chosen persons, were studied. In the first, individuals were born between 1940 and 1960, while in the second the time interval is 1961 – 1980. The data we receive contains only a set of two-dimensional vectors (face landmarks) obtained from the widely used, Dlib library **[9]**. The software provides 68 characteristic points in the face (landmarks), from which we select 26 landmarks based on their importance in the underlying osseous structure and stability with respect to changes in facial expression. Details may be found in the Methods section. In brief, we receive from a national database an anonymized pull of numbers, respecting in this way the confidentiality of image data and making possible a kind of epidemiological study.

From pairs of face points in a normalized image we compute distances. At the end, a vector of 24 distances comes up from each original photo. We show in (Figure 2) bottom panel the used face landmarks in a public image. Additional tests are applied to the set in order to detect, for example, deviations from frontality or any other artifact (See the Materials and Methods section for details).

The data is processed by means of PCA. The first 5 components are shown to account for nearly 90 % of data variance. We restrict ourselves to these 5 components and divide the PCA space in 15 cells corresponding to well defined groups of similar face characteristics. The results are shown in (Figure 3) in the form of a histogram comparing the two cohorts. The x-axis is a number labeling the cell, whereas the y-axis is the population density per 1000 inhabitants. That is, if the column height for a given cell is 300, there are 300 men with these facial characteristics per 1000 inhabitants.

We notice that there are measurable differences between cohorts in spite of the relatively small time lapse between them, 20 years. We checked by a kind of bootstrap analysis that in the cells with high differences error bars are small enough, thus differences are not artifacts. Our hypothesis is that the differences signal a change in the mixing dynamics inside the population after the event of Cuban revolution in 1959, with the abolishment of racial prejudices. As mentioned, these results are preliminary and should be further confirmed.

**2.2 Face morphometric profiles as early markers?**

A figure like (Figure 3), comparing the 60-80 cohort with data for Alzheimer or Parkinson patients born in the same time interval, for example, would indicate whether there is some population group with higher or lower risk for these diseases. The null hypothesis is that Alzheimer's disease (AD) patients are randomly distributed inside the population. Thus, the density of AD patients inside a given cell should be proportional to the cell density, with the same proportionality constant for all cells. Deviations in a given cell would indicate an increased (or decreased) risk. Of course, the number of AD patients should be high enough to get relevant statistics. Population-wide studies are needed. A preliminary study is already running in Cuba **[10]**. The risk for certain cancers could also be correlated to FM. On the other hand, as morphometric characteristics are more related to osseous structure than to expression or fat in face, indications may be used as early risk markers for the younger groups in the same morphometric groups.

It is worth discussing recent findings on the genetics of face and brain shape **[11]**. Observed correlations are only between FM and brain shape, not affecting behavioral-cognitive traits, in particular AD. One could rephrase these results in a simplified way as the absence of genetic correlations between phase morphometry and early AD.

However, AD is mostly a disease of the elderly **[12]**, essentially multifactorial. They are even hypothesis of infectious origin **[13]**. These multiple causes evolve against a given genetic background, which may accelerate or slow down progression towards AD. By measuring correlations between FM and AD we would point out predisposition to AD in later stages of life.

Consider, for example, the more limited set of 51 genes identified in **Ref. [14]** as involved in facial development. We list them in (Table 1). From them, 16 genes have been related to AD. We indicate in the Table the links to relevant literature.

In the same way, we indicate in the Table that 21 of these genes are related to cancer, according to Genecards **[15]**. An example is the Tumor Protein P63 gene (TP63), a member of the P53 family of transcription factors. At the mutational level there is not correlation to cancer. Probably, a statement like that of Ref. **[11]** may be formulated: No direct genetic correlation between FM and early cancer. However, cancer is also a multi-factorial disease of the elderly, and the baseline expression level of mutated TP63 could be a factor facilitating evolution towards prostate cancer, for example. Indeed, we identified TP63 among the 33 most important expression markers in prostate cancer **[16]**.

In order to check whether the expressions of the set of 51 genes play a role in defining the tumor state, we perform PCA calculations for TCGA expression data **[17]** in a set of tissues. Only the 51 genes in the set are used to conform the PCA matrix. Details on methods can be found elsewhere **[18]**. In general, this limited set is able to discriminate between a normal tissue and a tumor in many tissues, with low confusion matrices. We show in (Figure 4) an example of perfect discrimination in Glioblastoma. This seemly surprising result reinforces the idea that the baseline expressions of these genes could play a role in late cancer.

**3. Conclusions**

We reported preliminary results for the FM profile of the Cuban population and indicate that it could be used to find risk markers for groups inside the population for diseases such as Alzheimer's, Parkinson's and certain cancers. The idea behind this statement is the following. First, people inside a FM group share a considerable amount of genetic background **[19]**. Second, multifactorial, late diseases such as those mentioned above develop against the genetic background of each individual, thus that the background itself may be considered a risk factor. By comparing the density of Alzheimer patients with the population density one may, in principle, determine whether there are FM groups with predisposition to AD. As FM groups are based on osseous structure they could be used as early markers for younger groups with the same characteristics, allowing more detailed studies inside the group and early diagnosis.

**4. Materials and Methods**

**4.1. The dataset.** As mentioned, a dataset of random 200,000 facial images of Cuban men was processed by means of the DLIB software. The images correspond to two age cohorts - the first comprising men born between 1940 and 1960, and the second including men born between 1961 and 1980. Racial information was also available for each image, but not used.

The landmark localization model was unsuccessful in extracting facial landmarks for 1,968 images in the 40-60 years cohort and 470 images in the 61-80 years cohort. Combined, these landmark detection failures accounted for approximately 1% of the total image dataset. As such, the missing landmark data is not expected to substantially impact downstream analyses.

**4.2. Facial landmarks.** Facial landmark localization was performed using the 68-point face model from the DLIB library **[9]**. A representation of such points is shown in (Figure 2) top panel **[20]**. After extracting the landmark coordinates from each image, the landmarks were filtered to select only those representing relevant facial features that accurately reflect the underlying osseous structure. They are represented in (Figure 2) bottom panel in an image of the public UTKFace dataset (https://susanqq.github.io/UTKFace/, **[21])**. This accuracy is captured by landmarks that are most stable against variations in facial expression. Consider, for example, the analysis of eye landmarks. Point 38 would vary with the openness of the eye, while point 39 maintains its position despite such variations. Selecting stable points according to this criteria ensures important noise reduction and guarantees better reproducibility of results across different images of the same person.

Based on these stability criteria, the selected DLIB landmarks were: 0, 4, 7, 8, 9, 12, 16, 17, 19, 21, 22, 24, 26, 27, 30, 31, 33, 35, 36, 39, 42, 45, 48, 51, 54, 57, for a total of 26 points per image.

We received from a national database only the DLIB landmarks, with no reference to individual identity.

**4.3. Data filtering.** To ensure all facial images exhibit a frontal pose and neutral expression, two filtering criteria were imposed. First, frontal pose was quantified through a symmetry coefficient defined as: $s = \text{distance}(0,8) / \text{distance}(16,8)$, requiring that $0.90 < s < 1.10$, where distance(0,8) denotes the distance between landmarks 0 and 8, for example.

Second, mouth closure was assessed via a coefficient given by: $m = y66 / y62$, requiring that $m < 1.02$. In this case y62 and y66 correspond to the vertical coordinates of the central upper and lower lip landmarks 62 and 66. Tighter clustering of these landmarks indicates mouth closure. Notice that these landmarks are used only for filtering purposes.

**4.4. Building the distance matrix.** Rather than absolute facial landmark coordinates, the distances between landmarks are more informative. Two highly stable landmarks were chosen as reference points, with all other points defined by their normalized distances to these references. Specifically, the outer eye corners (points 36 and 45) served as the reference landmarks in this study, with the distance between them constituting the reference distance for a given image. Each remaining landmark distance was normalized to this image-specific reference distance, resulting in comparable measurements across the dataset.

To reduce dimensionality, symmetric and homologous landmark distances were consolidated by averaging. Symmetric distances were defined as those between landmarks along the mid-sagittal plane (points 8, 27, 30, 33, 51 and 57) and each of the references (e.g. $d_{8-36}$ and $d_{8-45}$). Equivalent distances were defined as those between symmetrical counterpart landmarks across the mid-sagittal facial plane and the reference landmark on the opposite side of the face (e.g. $d_{12-36}$ and $d_{4-45}$). Equivalent landmark pairs consisted of (0, 16), (4, 12), (7, 9), (17, 26), (19, 24), (21, 22), (31, 35), (39, 42) and (48, 54). After averaging of symmetric and equivalent distances, each facial image was characterized by a 24-dimensional distance vector.

**4.5. Density map.** To reduce dimensionality and decorrelate the data, principal component analysis (PCA) was performed on the distance data. The facial data samples were then projected onto the first 5 principal components, which account for nearly 90% of the variance. The vectors defining the principal components for the 40-60 years age cohort are used in the 61-80 cohort as well in order to compare them. The space was divided into equal-width intervals. The number of subdivisions in each component is taken roughly proportional to its variance. Specifically, the first principal component was partitioned into 5 intervals and the second into 3 intervals, resulting in a total of 15 cells partitioning the PC1 vs PC2 plane, and thus the whole space. Samples are count in each cell in order to to determine the incidence per 1000 individuals and construct the 2D density map.

**Acknowledgements**

Authors are grateful to Yasser Perera and Gabriel Gil for useful discussions. The authors acknowledge the Cuban Agency for Nuclear Energy and Advanced Technologies (AENTA) and the Office of External Activities of the Abdus Salam Centre for Theoretical Physics (ICTP) for support.

**Compliance with ethical standards**

**Conflict of Interest:** Authors declare that they have no conflict of interest.

**Funding:** The work was self-funded by the authors.

**Ethical approval:** This article does not contain any studies with human participants or animals performed by any of the authors.

**Informed consent:** For this type of study, formal consent is not required.

**Author contribution**

Roberto Herrero: Data curation and processing, data visualization, formal analysis, investigation, methodology and validation.

Yoanna Martinez: Data curation and processing, investigation, methodology and validation.

Heydi Mendez: Data curation and processing, supervision, writing-review and editing.

Joan Nieves: Data processing and visualization, investigation, methodology and validation.

Augusto Gonzalez: Conceptualization, formal analysis, methodology, supervision, writing original draft, review and editing.

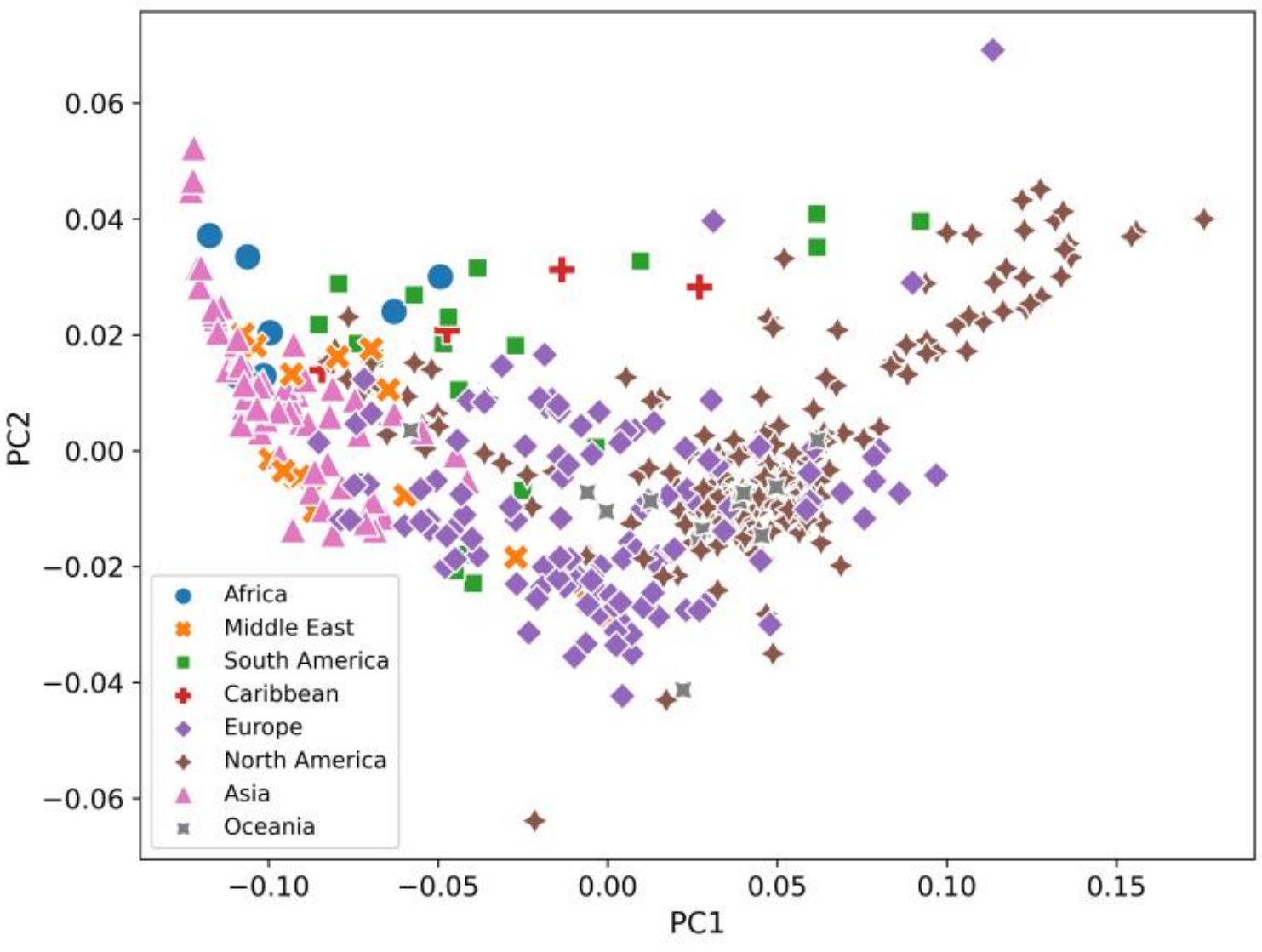

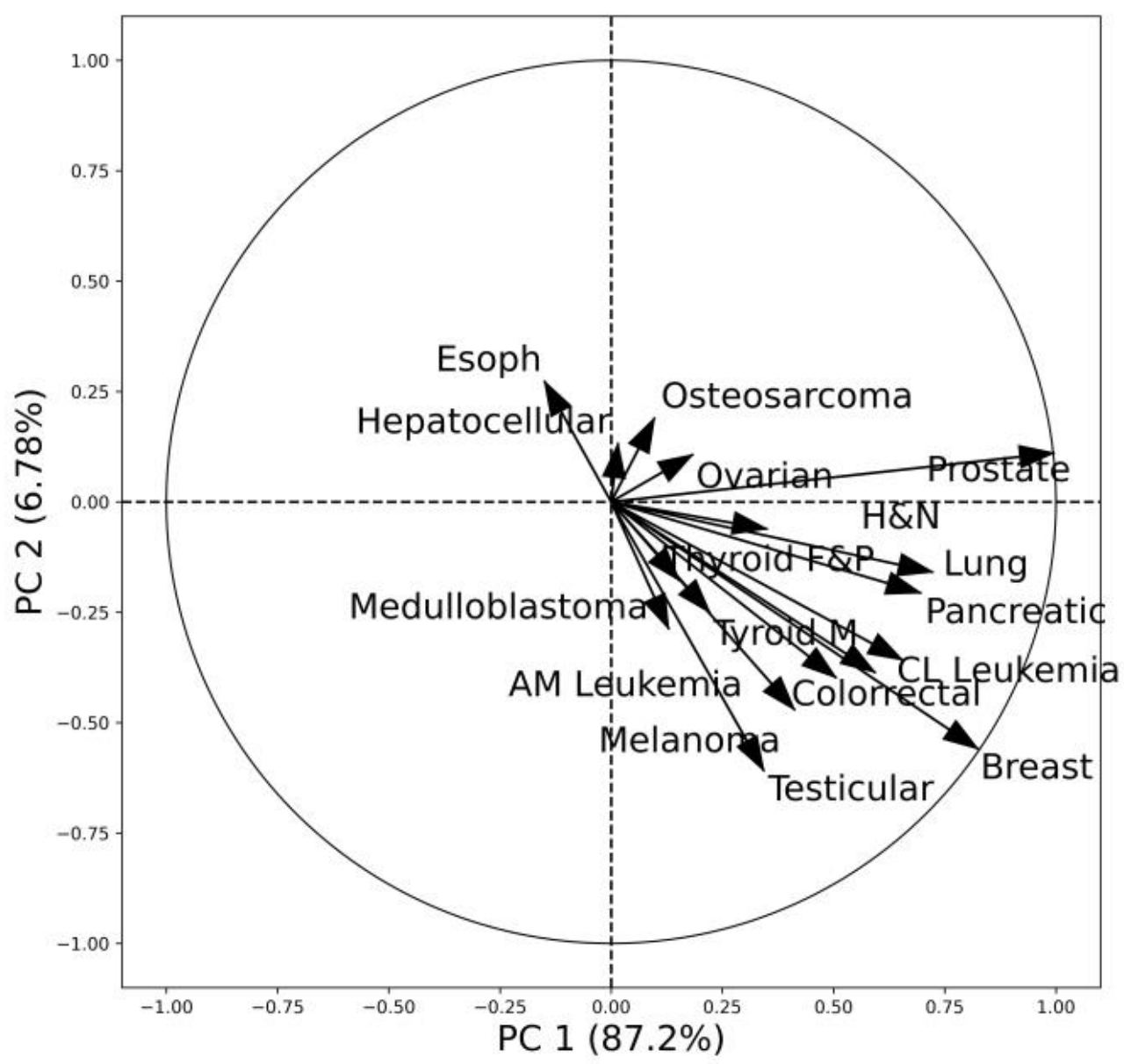

**Fig. 1. Top:** Principal Component Analysis of world cancer risk data. Ethnic and cultural groups exhibit distinctive patterns. **Bottom:** Component variances and directions of maximal increase in the risk for different cancers.

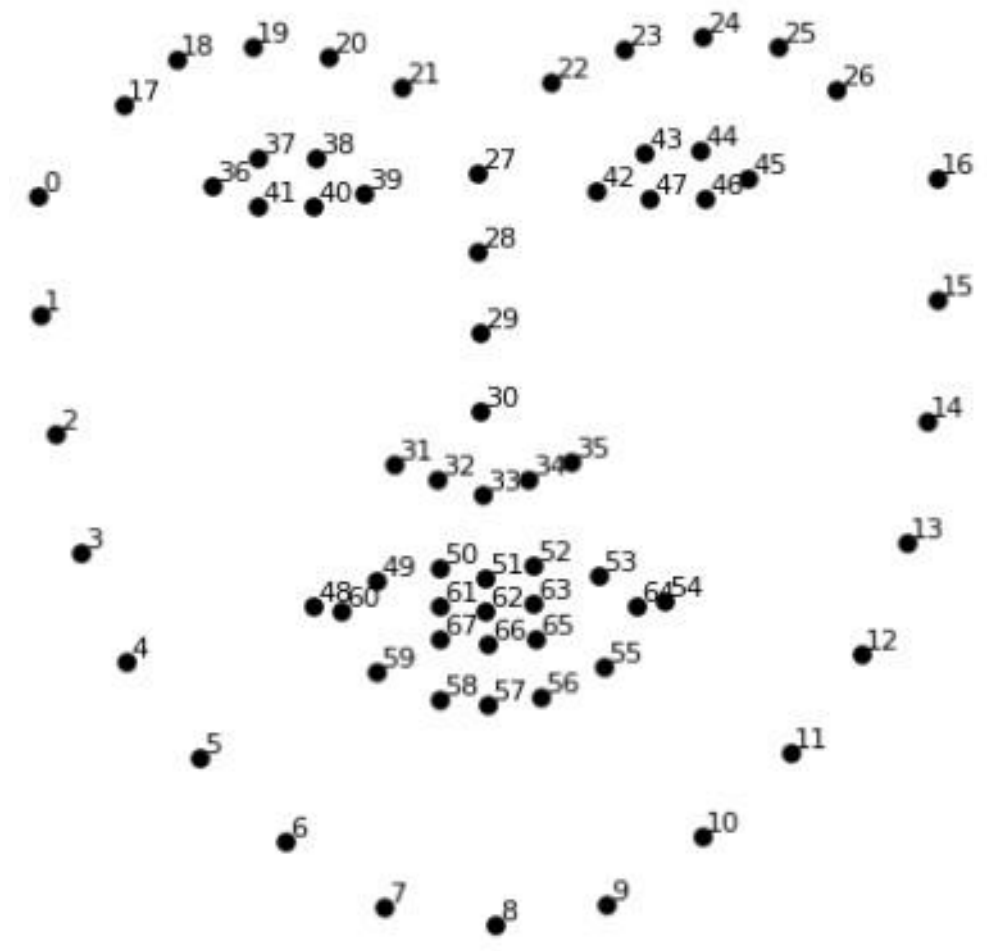

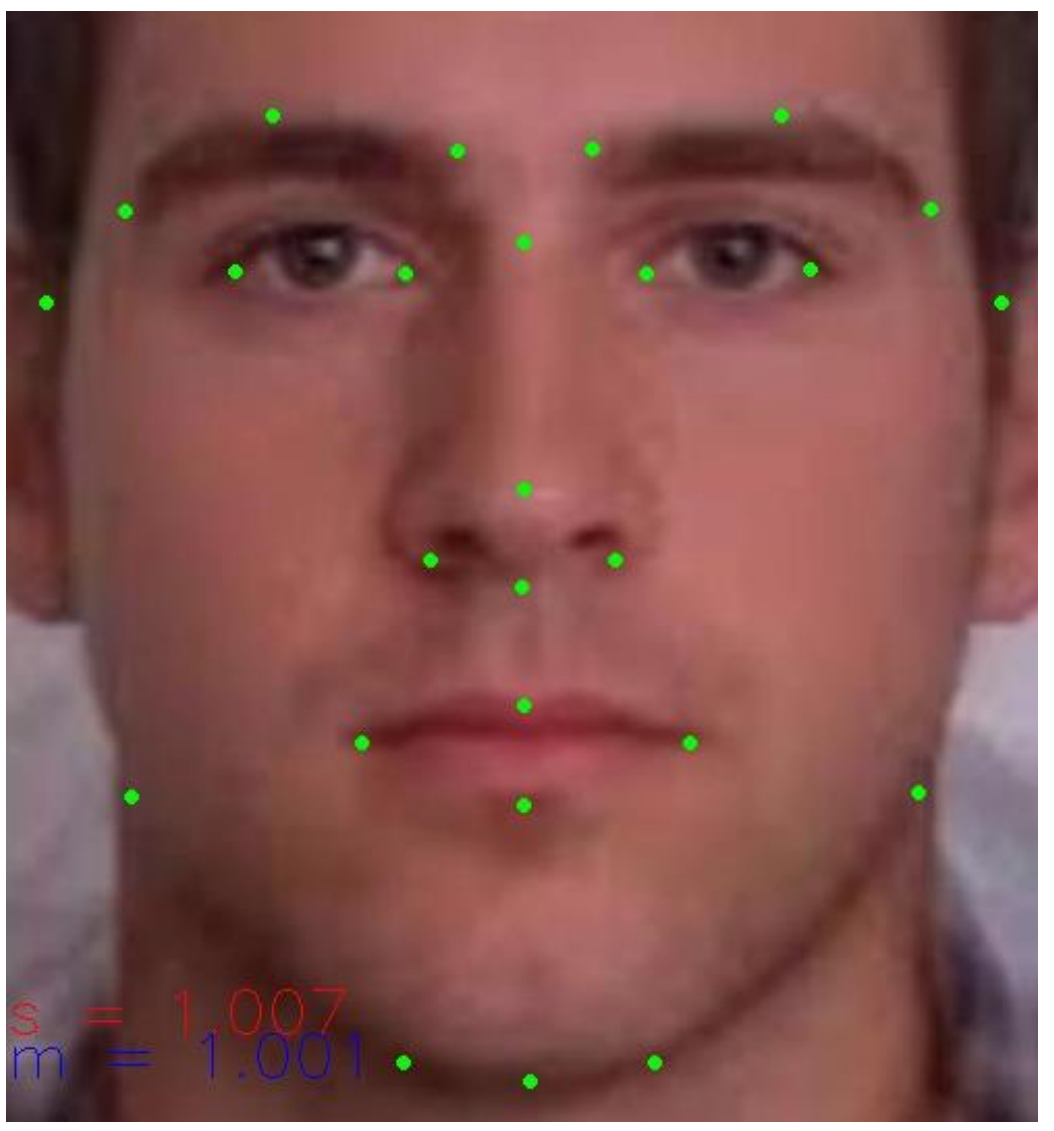

**Fig. 2. Top:** Face landmarks coming from the DLIB software. **Bottom:** The selected landmarks, which capture osseous structure and are relatively independent of face expression (green points).

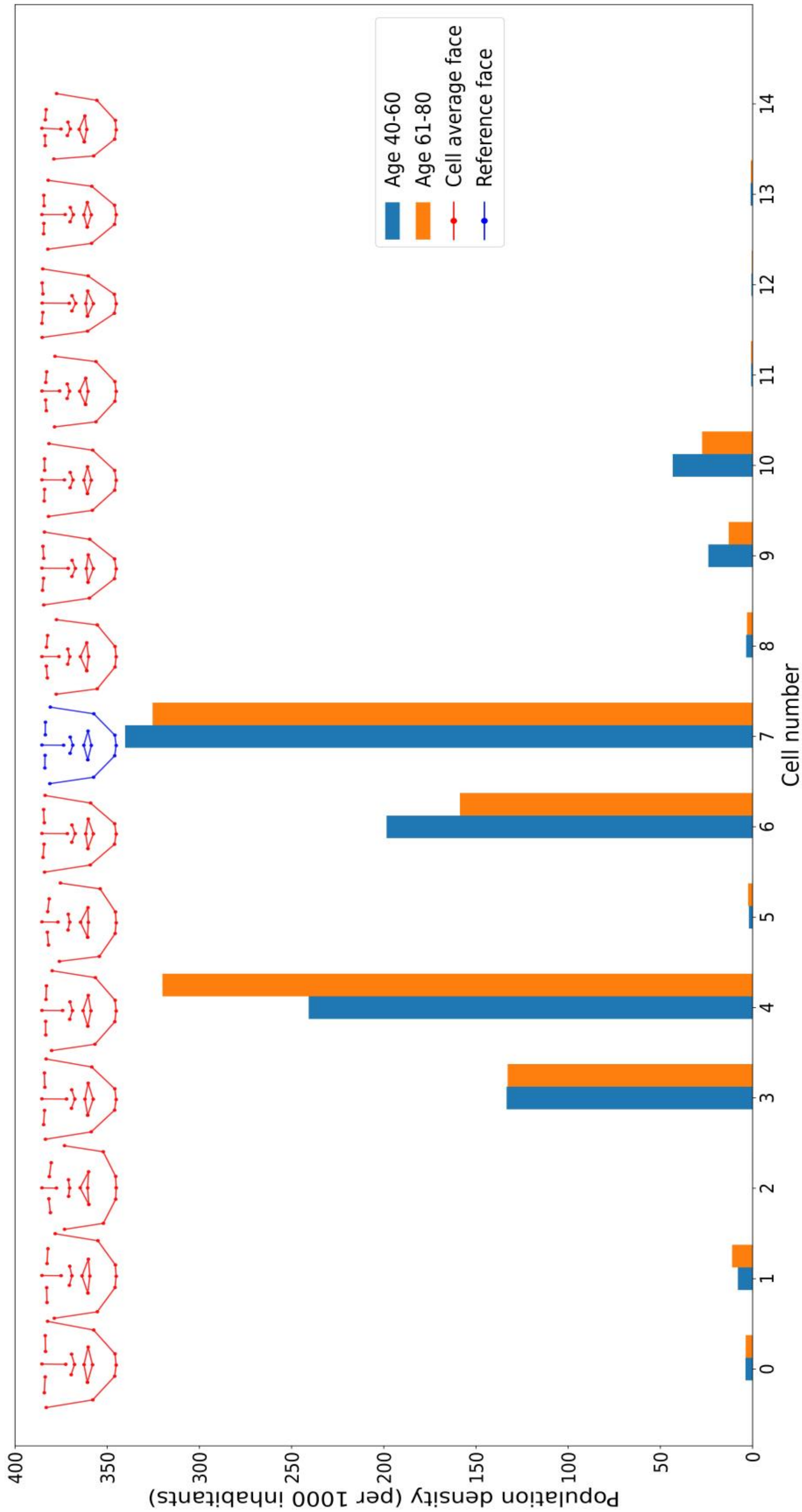

**Fig. 3.** Preliminary results for the face morphometry profile of the Cuban population. Two cohorts are compared, one from men born between 1940 and 1960, and the second from men born between 1961 and 1980. A schematics of face characteristics is shown on top.

Role in cancer and AD of genes involved in facial development

| Gene | Alias | Role in facial development (Frontiers in Genetics) | Role in cancer (Genecards) | Role in AD | Ref. for AD |
|---|---|---|---|---|---|
| ACAD9 | Acyl-CoA Dehydrogenase Family Member 9 | Philtrum | Not reported | Not reported | |
| ALX3 | ALX Homeobox 3 | Eye width | Neuroblastoma | Not reported | |
| ASPM | Assembly Factor For Spindle Microtubules | Chin prominence | Cancer related | Brain Size, AD protection | https://www.sciencedirect.com/science/article/pii/S2352873719300642 |
| CASC17 | Cancer Susceptibility Candidate 17 (Non-Protein Coding) | Nose prominence | No? | Not reported | |
| CHD8 | Chromodomain Helicase DNA Binding Protein 8 | Nose prominence | Cancer related | Not reported | |
| COL17A1 | Collagen Type XVII Alpha 1 Chain | Eye width and depth | Cancer related | Not reported | |
| SMIM23 (C5orf50) | Small Integral Membrane Protein 23 | Nasion, eyes and zygoma prominence | Not reported | Not reported | |
| DCHS2 | Dachsous Cadherin-Related 2 | Ala aperture, Nose prominence | Not reported | AD Related, age of onset | https://www.nature.com/articles/mp2011135 |
| DHX35 | DEAH-Box Helicase 35 | Nose width | Not reported | AD Related (Not clear) | https://www.frontiersin.org/articles/10.3389/fpsyt.2019.00843/full |
| DLX6 | Distal-Less Homeobox 6 | Chin prominence | Not reported | | |
| DVL3 | Dishevelled Segment Polarity Protein 3 | Nose bridge | Cancer related | Dysregulated in AD | https://www.nature.com/articles/s41598-019-54782-y |
| DYNC1LI1 | Dynein Cytoplasmic 1 Light Intermediate Chain | Chin prominence | Not reported | Not reported | |
| EDAR | Ectodysplasin A Receptor | Chin prominence and shape | Breast cancer | Not reported | |
| EPHB3 | EPH Receptor B3 | Nose bridge | Cancer related | Not reported | |
| EYA4 | EYA Transcriptional Coactivator And Phosphatase 4 | Forehead | Not reported | Not reported | |
| FOXA1 | Forkhead Box A1 | Face width | Cancer related | AD related | https://www.sciencedirect.com/science/article/abs/pii/S0888754321003189 |
| FREM1 | FRAS1 Related Extracellular Matrix 1 | Lip (upper) | Not reported | Not reported | |
| GLI3 | GLI Family Zinc Finger 3 | Nose width | Basal cell, Brain and Colorectal cancer | Language dysfunction in AD | https://www.ncbi.nlm.nih.gov/pmc/articles/PMC5583024/ |
| GNAI3 | G Protein Subunit Alpha I3 | Eye width | Not reported | AD possibly related | https://www.biorxiv.org/content/10.1101/322503v1.abstract |
| GSTM2 | Glutathione S-Transferase Mu 2 | Eye width | Hepatocell. Carcinoma | Not reported | |
| HDAC8 | Histone Deacetylase 8 | Eye width | Not reported | Not reported | |
| HOXD1 | Homeobox D1 | Eye shape, Lip prominence | Not reported | Not reported | |
| KCTD15 | Potassium Channel Tetramerization Domain Containing 15 | Nose prominence | Not reported | Not reported | |
| MAFB | MAF BZIP Transcription Factor B | Face width | Cancer related | Expressed Transcription Factors in AD | https://www.frontiersin.org/articles/10.3389/fnagi.2022.881488/full |
| MBTPS1 | Membrane Bound Transcription Factor Peptidase, Site 1 | Upper facial profile height | Not reported | Not reported | |
| MIPOL1 | Mirror-Image Polydactyly 1 | Face width | Bronchus adenoma | Not reported | |
| MTX2 | Metaxin 2 | Eye shape | Not reported | Not reported | |
| OSR1 | Odd-Skipped Related Transcription Factor 1 | Face height/depth | Not reported | Not reported | |
| PABPC1 | Poly(A) Binding Protein Cytoplasmic 1 Pseudogene 1 | Eye width | Not reported | AD Related | https://onlinelibrary.wiley.com/doi/full/10.1111/cns.13117 |
| PRKN | Parkinson Protein 2, E3 Ubiquitin Protein Ligase | Face height | Ovarian, Lung cancer | Mitophagy, AD related | https://alz-journals.onlinelibrary.wiley.com/doi/full/10.1002/alz.12198 |
| PAX1 | Paired Box 1 | Nose width | Ovarian, Cervical cancer | Not reported | |
| PAX3 | Paired Box 3 | Eye width, Nasion prominence, Nose width | Not reported | AD Related | https://content.iospress.com/articles/journal-of-alzheimers-disease/jad140729 |
| PAX9 | Paired Box 9 | Face width | Lung cancer | Not reported | |
| PCDH15 | Protocadherin Related 15 | Upper facial profile prominence | Not reported | AD Related | https://link.springer.com/article/10.1007/s10048-010-0234-9 |
| PDE8A | Phosphodiesterase 8A | Allometry | Not reported | Not reported | |
| PKDCC | Protein Kinase Domain Containing, Cytoplasmic | Mental fold | Not reported | Not reported | |
| PRDM16 | PR-Domain Zinc Finger Protein 16 | Nose prominence and width | Cancer related | Not reported | |
| RAB7A | RAB7A, Member RAS Oncogene Family | Philtrum | No? | AD Related | https://pubmed.ncbi.nlm.nih.gov/26768426/ |
| RPS12 | Ribosomal Protein S12 | Forehead | Not reported | Not reported | |
| RUNX2 | RUNX Family Transcription Factor 2 | Nose width | Prostate cancer, Leukemia | Not reported | |
| SCHIP1 | Schwannomin Interacting Protein 1 | Centroid size | Not reported | Not reported | |
| SLC25A2 | Solute Carrier Family 25 Member 2 | Face width | Not reported | Not reported | |
| SOX9 | SRY (Sex-Determining Region Y)-Box 9 Protein | Nose prominence and width | Cancer related | Potenial AD related | https://www.sciencedirect.com/science/article/abs/pii/S0006899316304346 |
| SUPT3H | Suppressor Of Ty 3 Homolog | Nose width | Hepatic adenomas | Not reported | |
| TBX15 | T-Box Transcription Factor 15 | Forehead | Not reported | AD related | https://genomemedicine.biomedcentral.com/articles/10.1186/s13073-015-0258-8 |
| TMEM163 | Transmembrane Protein 163 | Eye width and depth | Not reported | Not reported | |
| TP63 | Tumor Protein P63 | Eye width | Cancer related | Potentially AD related | https://www.sciencedirect.com/science/article/abs/pii/S1053811910000649 |
| TRPC6 | Transient Receptor Potential Cation Channel Subfamily C Member 6 | Facial depth | Colorectal cancer | AD related | https://www.ncbi.nlm.nih.gov/pmc/articles/PMC7816687/ |
| WDR27 | WD Repeat Domain 27 | Eye shape | Not reported | Not reported | |
| WDR35 | WD Repeat Domain 35 | Face height/depth | Not reported | Not reported | |

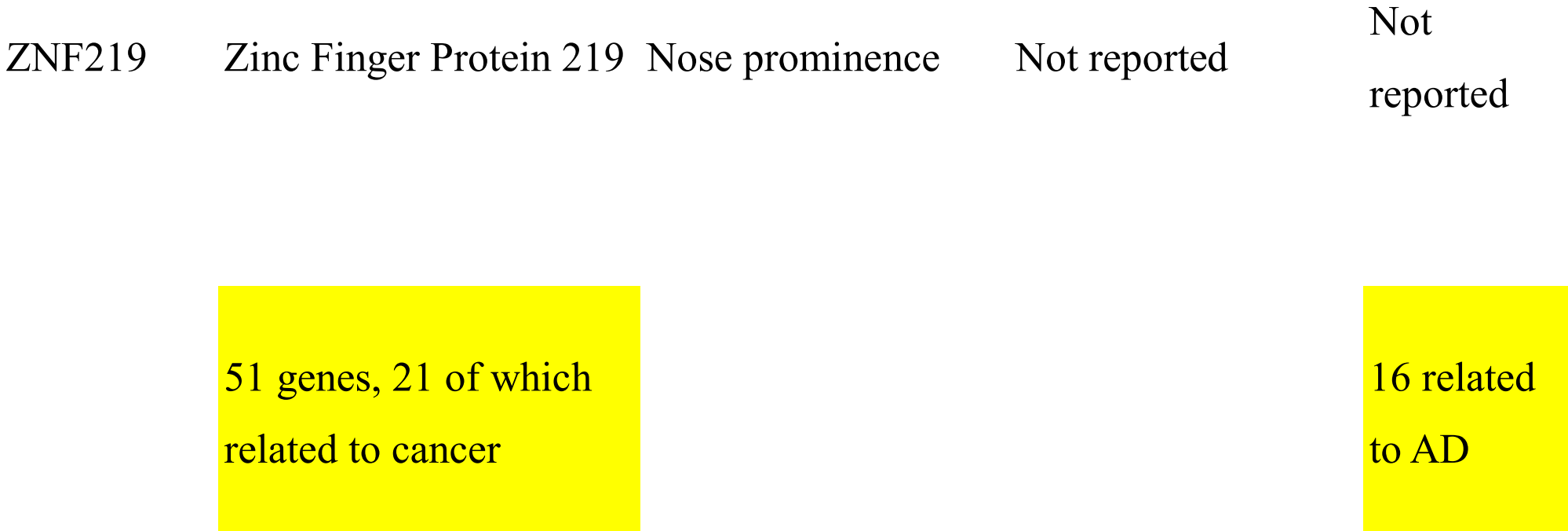

| ZNF219 | Zinc Finger Protein 219 | Nose prominence | Not reported | Not reported |
|---|---|---|---|---|
| | 51 genes, 21 of which related to cancer | | | 16 related to AD |

**Table I.** The 51 genes involved in facial development, according to **Ref. [14]**. We indicated whether they have been associated to cancer (Genecards) or to AD in the literature.

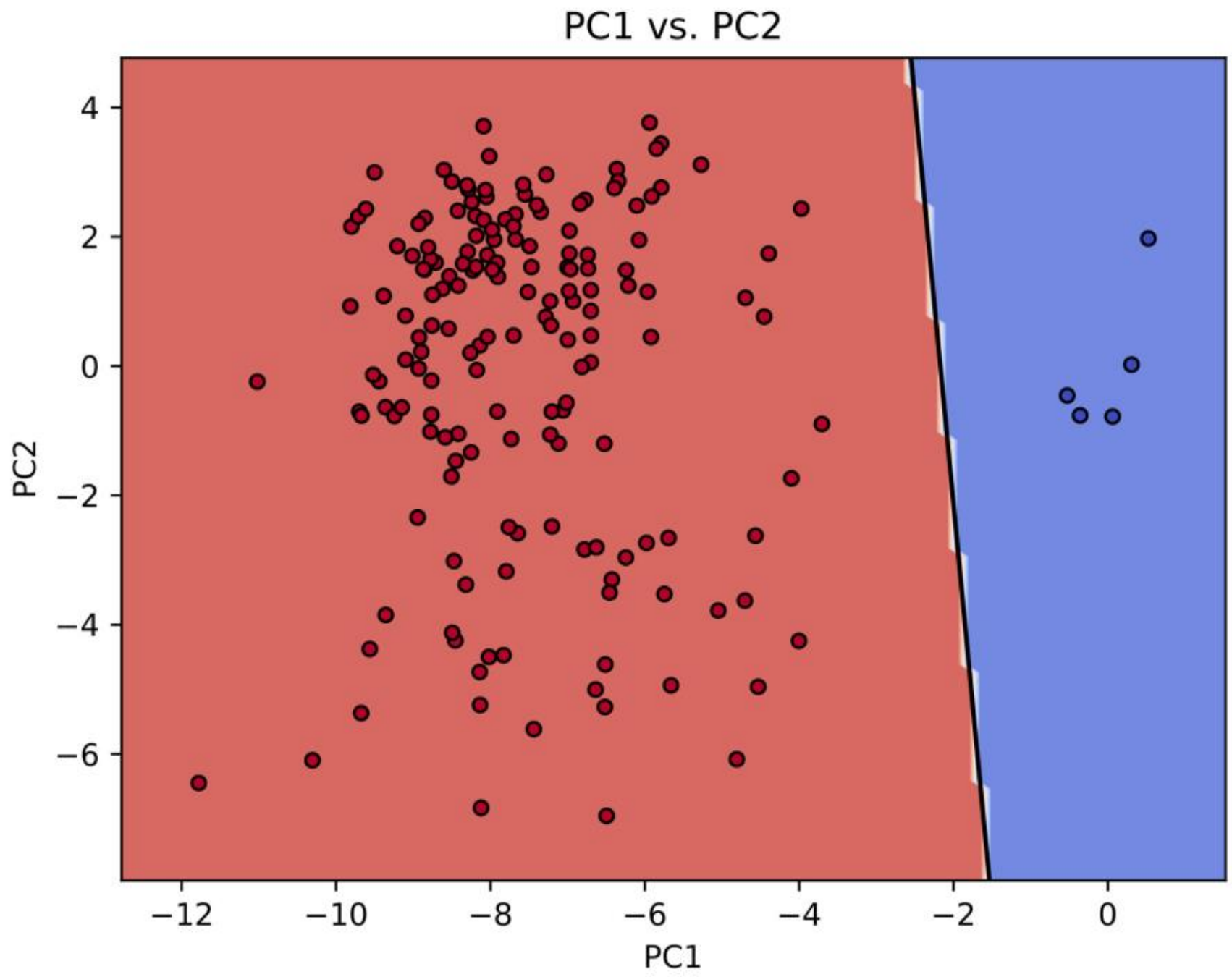

**Fig. 4.** Reduced PCA of TCGA expression data for Glioblastoma. Only the 51 genes of **Ref. [14]** are used to conform the PCA matrix.  Normal tissue (blue) and tumor data (red) are perfectly separated.